\documentclass[%
preprint,
superscriptaddress,
 amsmath,amssymb,
 aps,
]{revtex4-1}
\usepackage{amsmath}
\usepackage{graphicx}
\usepackage{hyperref}
\usepackage{siunitx}
\usepackage{lineno}

\usepackage{graphicx}
\usepackage{dcolumn}
\usepackage{bm}
\usepackage{verbatim}   
\usepackage{color}
\usepackage[normalem]{ulem}

\begin{document}


\title{Semiclassical description of Intermolecular Coulombic Electron Capture in solutions}

\author{Nicolas Sisourat}
\affiliation{
	Sorbonne Universit\'{e}, CNRS, Laboratoire de Chimie Physique Mati\`{e}re et Rayonnement, UMR 7614, F-75005 Paris, France
   }
\date{\today}
	\begin{abstract}
	In this work, we present a semiclassical approach to model Intermolecular Coulombic Electron Capture (ICEC) in aqueous solutions using molecular dynamics simulations with OpenMM.
	We investigate the behavior of an excess electron in the presence of cations (Fe$^{3+}$) in water, focusing on the influence of electron energy and cation concentration on the ICEC quantum yield.
	Our simulations reveal that the ICEC quantum yield approaches unity at higher concentrations and initial electron energies, while it decreases at lower concentrations due to electron energy loss before reaching the cation.
\end{abstract}
\maketitle

\section{Introduction}
When high-energy particles or photons interact with living cells, they can eject photoelectrons from water molecules within the cell. These ejected electrons initiate a cascade of chemical reactions and energy transfer processes, leading to significant biological effects. The dynamics of electron scattering in liquid water are of paramount importance, not only in the field of condensed matter physics but also in radiation chemistry and radiation biology. Understanding these processes is essential for elucidating the mechanisms of radiation-induced damage, which can have profound implications for both therapeutic applications, such as radiation therapy, and for assessing the risks associated with radiation exposure~\cite{Nikjoo2012}.

Intermolecular Coulombic Electron Capture (ICEC) is a fundamental energy transfer process where an electron is captured by a cation while the excess energy is transferred to a neighboring species which is thus ionized~\cite{gokhberg_environment_2009,gokhberg_interatomic_2010,bande_interatomic_2023}. The cation is thus reduced while a surrounding molecule is oxidized. 

ICEC has been extensively investigated across a range of systems, including atomic pairs and molecular clusters~\cite{Sisourat18,Molle21,Molle23}. For instance, systems such as H$^+$ + H$_2$O and H + H$_2$O$^+$ have been analyzed using ab initio methods to elucidate the underlying mechanisms and cross sections of the process~\cite{Graves24}. Recent theoretical advancements have integrated vibrational motion and temperature dependence into models of ICEC cross sections, revealing that these factors can modestly reduce the efficiency of ICEC while maintaining its dominance over competing processes such as photorecombination~\cite{Jahr25}.
Additionally, ICEC has been explored in quantum dot systems, where it is driven by long-range electron correlations. Research in this area has demonstrated that ICEC can facilitate the transition between the capture of free-moving electrons and the release of trapped electrons in neighboring regions, such as in nanowire-embedded quantum-dot pairs~\cite{Pont13,Pont16,Molle19}. It is also noteworthy that several related processes and extensions of ICEC have been proposed, further expanding the scope and potential applications~\cite{Muller10,Jacob19,Remme23,Ellerbrock25,Drennhaus25}.

While understanding ICEC in aqueous solutions is crucial for various applications, including radiation therapy and environmental chemistry, it remains largely unexplored. In this study, we employ a semiclassical approach to simulate ICEC in water, using the OpenMM molecular dynamics engine. We explore the dynamics of an excess electron in the presence of cations and analyze the factors influencing the ICEC quantum yield. In particular, we focus on  ferric (Fe$^{3+}$) ions which are crucial for numerous biological processes. 

\section{Methods and Computational details}

The system consists of a box of water molecules with an excess electron and one or two cations (depending on the molar concentration of the cation). The water molecules are modeled using the TIP3P water model \cite{Jorgensen83}, which is a rigid, non-polarizable model with three interaction sites.

The interactions between water molecules are described by the AMBER force field \cite{Case2005}. The electron and the cations are treated as classical particles. The interactions between them and  with the water molecules are modeled using a custom non-bonded force that includes Coulombic and repulsive terms.

Periodic boundary conditions are applied to simulate an infinite system. The Particle Mesh Ewald method is used for long-range electrostatic interactions, and a cutoff is applied for van der Waals interactions.

The simulations are performed using the Langevin integrator to maintain a constant temperature. The friction coefficient is set to zero. However, we apply a damping force to the electron to mimic the electron energy loss in water. The stopping power cross sections are taken from~\cite{Castillo21}. The dynamics reproduces well the energy attenuation length as in~\cite{Suzuki14}.

A water box is created using the TIP3P water model. The size of the box is typically between 1.0 and 4.0 nanometers on each side, containing between approximately 100-2000 water molecules (C=1.6-0.025 mol/L (also noted M hereafter) ). The electron and the cation(s) are placed randomly within the box (a snapshot of the simulation box is shown in Fig. 1).

\begin{figure}[h]
	\centering
	\includegraphics[width=0.4\textwidth]{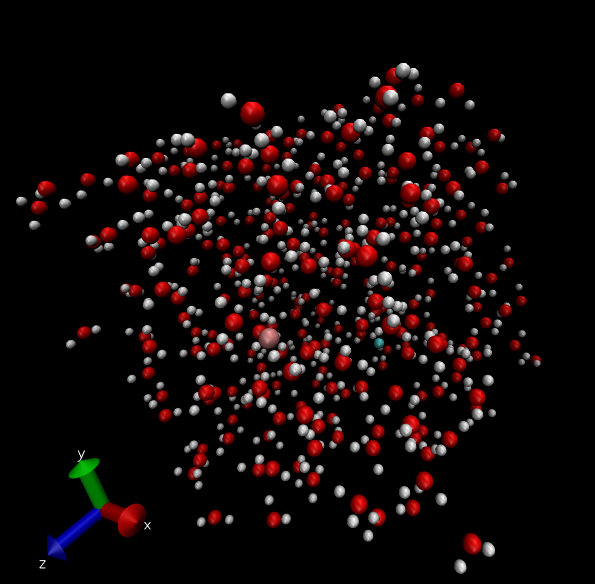}
	\caption{Snapshot of the simulation box: O are in red, H in white, Fe in pink and e in blue.}
	\label{fig:fig1}
\end{figure}

The interaction between the particles is described by a custom non-bonded force with the following energy expression:
\[ E = \frac{1}{4\pi\epsilon_0\epsilon_r} \frac{q_1 q_2}{r} + A \cdot e^{-r/\rho} \]
where \(q_1\) and \(q_2\) are the charges of the particles, \(r\) is the distance between them, \(A\) is the repulsion strength (\SI{1000}{\kilo\joule\per\mole}), and \(\rho\) is the repulsion range (\SI{0.03}{\nano\meter}). The vacuum  relative permittivity is noted $\epsilon_0$ and $\epsilon_r$ is dielectric constant of water.

The simulations are performed with the following parameters: the temperature is fixed at \SI{300}{\kelvin}, the integration timestep is chosen as \SI{2}{\femto\second} and the non-bonded cutoff is set at \SI{1.0}{\nano\meter}.

The simulation protocol involves the following steps:
\begin{enumerate}
	\item Energy minimization to remove bad contacts.
	\item Equilibration run to stabilize the system.
	\item Production run to collect data.
\end{enumerate}

The trajectory of all particles is saved at regular intervals. The kinetic energy of the electron is recorded at each step to analyze its behavior over time.

The trajectory is analyzed to calculate the distance between the electron and the cation r$_i$(t) ($i$=1 or 2). These distances are used to compute the probabilities that ICEC takes place at time t. The latter are chosen as follows:

The ICEC cross section, \(\sigma_{ICEC}\), represents the effective area for an interaction between the electron and a cation. Consider a particle moving through a medium with a number density of targets, \( n \). The number of interactions per unit time in a small volume \(\Delta V = A \Delta x\) is given by:
\[
\text{Number of interactions} = \sigma_{ICEC} \cdot n \cdot \Delta x \cdot \text{Number of incident particles}.
\]
Accordingly, the probability that a single incident electron interacts while traveling a small distance \(\Delta x\) is:
\[
P_{\text{ICEC}} = \sigma_{ICEC} \cdot n \cdot \Delta x.
\]
For a finite distance \( x \), the probability that a particle does not interact over the distance \( x \) is:
\[
P_{\text{no ICEC}}(x) = e^{-\sigma_{ICEC} n x}.
\]
This is derived by considering the product of the probabilities of not interacting in each infinitesimal segment \( dx \):
\[
P_{\text{no ICEC}}(x) = \prod_{i=1}^{N} (1 - \sigma_{ICEC} n \, dx) \approx e^{-\sigma_{ICEC} n x}.
\]
Thus, the probability that an interaction does occur over distance \( x \) is:
\[
P_{\text{ICEC}}(x) = 1 - e^{-\sigma_{ICEC} n x}.
\]
In our simulations, we work with time steps, \(\Delta t\). The distance traveled in time \(\Delta t\) is \(\Delta x = v \Delta t\), where \( v \) is the relative velocity which is assumed to be constant within  \(\Delta t\). Substituting \(\Delta x\) into the probability formula gives:
\[
P_{\text{ICEC}}(t) = 1 - e^{-\sigma_{ICEC} n v \Delta t}.
\]
Finally, in our case $n$ depends on time and we assume $n(t) = \frac{1}{r_i(t)^3}$, thus we obtain
\[
P_{\text{ICEC}}(t) = 1 - e^{-\sigma_{ICEC} v \Delta t r_i^{-3}}.
\]

The ICEC cross section ($\sigma_{ICEC}(E_e)$) is estimated using the model developed in~\cite{Senk24}. At every timestep, a random number is drawn from uniform distribution (between 0 and 1). If $P_{\text{ICEC}}(t)$ is larger than this random number, we assume ICEC took place, provided that the electron has enough energy to trigger ICEC.

The ICEC quantum yield is obtained as
\[
	\Phi = N_{ICEC}/N_{tot}
\]
where $N_{tot}$, $N_{ICEC}$ are the total number of trajectories and the number of trajectories that undergo ICEC, respectively. We have performed 100 trajectories for each electron energy and molar concentration.

The simulations are performed using OpenMM \cite{Eastman2017}. The analysis is performed using Python with libraries such as NumPy, SciPy, and Matplotlib.

\section{Results}

Fig. 2 shows the distance between the electron and the Fe$^{3+}$ cation as a function of time in a solution of 0.2 mol/L (that corresponds to a box of 2x2x2 nm$^3$ and one cation or a box of 2.5x2.5x2.5 nm$^3$ and two cations; both simulations give similar results). The energy of the electron as a function of time is also shown.

As the electron scatters through the solvent (the cation moves much less), the distance between the electron and the cation oscillates. ICEC is most likely to take place at the shortest distances. However, the electron kinetic energy decreases fast due to inelastic collisions with the solvent molecules (modeled by a damping force; see Methods and Computational details). After less than 5fs, ICEC is energetically forbidden.

\vspace{1cm}

\begin{figure}[h]
    \centering
    \includegraphics[width=0.5\textwidth]{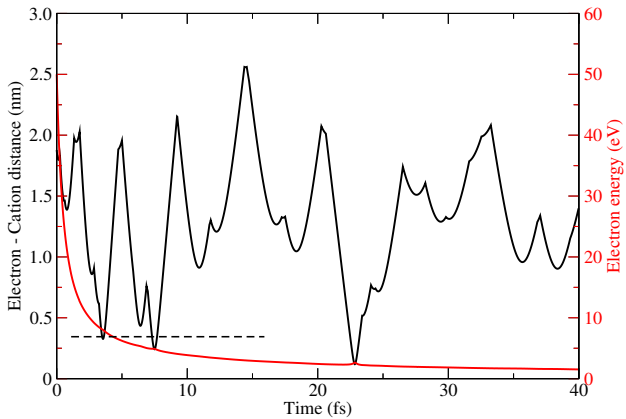}
    \caption{Distance (in nm) between the electron and the cation (in black) and electron kinetic energy (in red) as function of time for a typical trajectory at C=0.2mol/L. The dashed line indicates the ICEC threshold (i.e. below this line ICEC is not energetically possible).}
    \label{fig:fig2}
\end{figure}

Our simulations comprised 100 independent trajectories for each combination of cation concentration and initial electron kinetic energy. The resulting ICEC quantum yields are presented in Fig. 3 for three distinct initial electron energies (10 eV, 50 eV, and 100 eV) across a range of cation concentrations. At high cation concentrations, the ICEC quantum yield approaches unity. This indicates that the probability of an electron encountering and being captured by a cation is significantly increased when cations are more abundant, minimizing the likelihood of energy loss before capture.
Conversely, at lower cation concentrations, the electron is more likely to lose a substantial portion of its kinetic energy through inelastic collisions with solvent molecules before reaching a cation. This energy dissipation reduces the probability of ICEC, thereby lowering the quantum yield. Consequently, the ICEC quantum yield exhibits a marked decrease as the cation concentration is reduced.
To quantitatively describe the relationship between the ICEC quantum yield and cation concentration, we fitted the simulation results with a quadratic function. This fitting suggests that, within the range of concentrations considered, the ICEC quantum yield scales quadratically with cation concentration.

It should be noted that the plot in Fig. 3 can be obtained experimentally by measuring for example the UV absorption spectrum of the solution: ICEC leads to Fe$^{2+}$ which absorbs at different wavelength than Fe$^{3+}$~\cite{Fontana07}. One can irradiate a solution of pure Fe$^{3+}$ with electrons or with high-energy photons which create a swarm of electrons by photoionization of the solvent molecules. These electrons can undergo ICEC, reduce the concentration of Fe$^{3+}$ and increase that of Fe$^{2+}$.

\vspace{1cm}

\begin{figure}[h]
    \centering
    \includegraphics[width=0.5\textwidth]{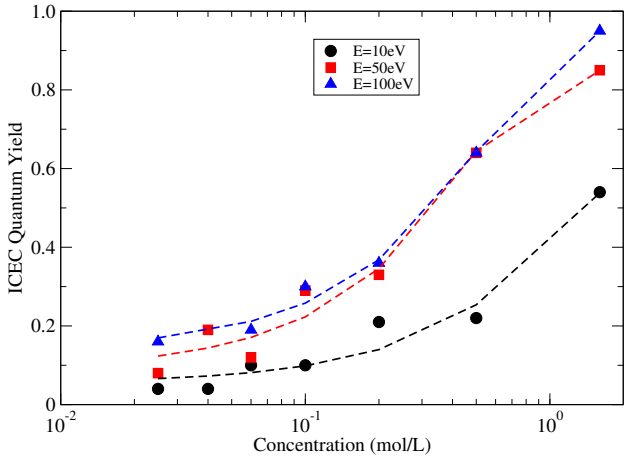}
    \caption{ICEC quantum yield as function of the cation concentration. The results from the simulations (dots) were fitted with a quadratic function (lines); see text.}
    \label{fig:fig3}
\end{figure}

An important quantity for radiation chemistry is the electron kinetic energy since the latter directly influences the physical, chemical, and biological effects of ionizing radiation. Figs. 4 shows the emitted ICEC electron energy as function of time for C=0.2 mol/L. At short time, the ICEC electron has fairly high energies. It corresponds to trajectories for which the electron reaches the cation quickly and thus do not significantly lose energies. At longer time, more electrons reach the cation but with lower energy. Note that such results can be obtained with pump-probe spectroscopy using for example, a fs pump pulse of high energy photons creating electrons and triggering ICEC and a fs probe UV pulse which allows to measure the absorption spectra at different delay time after the pump pulse. It should also be mentioned that the emitted ICEC electron can lose energy due to inelastic collisions leading to a solvated electron or can undergo another ICEC process if energy is high enough. This dynamics is not considered in this work.   

\vspace{1cm}

\begin{figure}[h]
    \centering
    \includegraphics[width=0.5\textwidth]{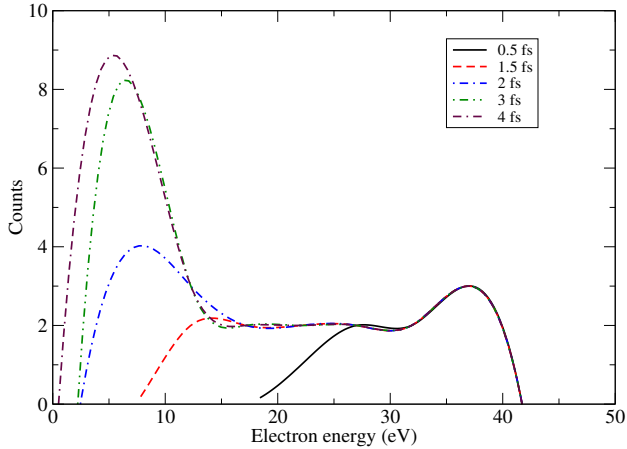}
    \caption{ICEC electron energy distribution for different times for C=0.2mol/L and the initial electron energy of 50 eV.}
    \label{fig:fig4}
\end{figure}

\begin{figure}[h]
	\centering
	\includegraphics[width=0.5\textwidth]{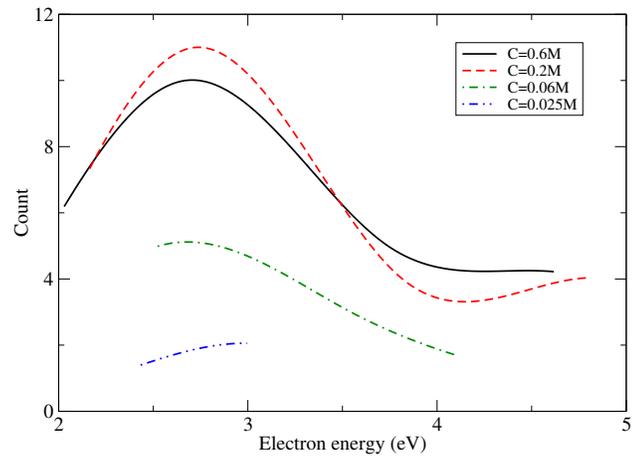}
	\vspace{1cm}
	
	\includegraphics[width=0.5\textwidth]{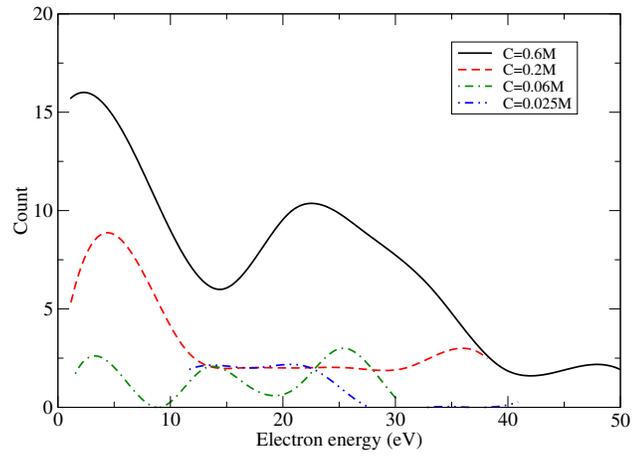}
	\vspace{1cm}
	
	\includegraphics[width=0.5\textwidth]{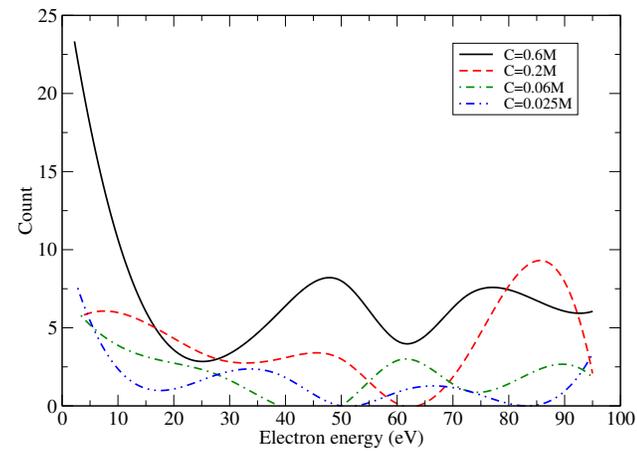}
	\caption{Total ICEC electron energy distribution for different concentrations. From top to bottom: initial electron energy = 10, 50 and 100 eV.}
	\label{fig:fig5}
\end{figure}

Finally, we show the total emitted ICEC electron energy for different concentrations and 3 initial kinetic energies in Fig. 5. The comparison within each panel shows that at lower concentration the ICEC electron is emitted at longer times with thus lower kinetic energy. Comparing the panels, we see that for lower initial kinetic energy the ICEC electron is relatively slower compared to initial electron.

\section{Conclusions}

In this study, we have presented a semiclassical approach to model Intermolecular Coulombic Electron Capture (ICEC) in aqueous solutions using molecular dynamics simulations with OpenMM. Our focus was on understanding the behavior of an excess electron in the presence of Fe$^{3+}$ cations in water, with particular attention to the influence of electron kinetic energy and cation concentration on the ICEC quantum yield.

Our simulations revealed that the ICEC quantum yield is highly dependent on both the initial kinetic energy of the electron and the concentration of the cations. At higher concentrations, the ICEC quantum yield approaches unity, indicating that the electron is more likely to encounter and be captured by a cation before losing significant energy. Conversely, at lower concentrations, the electron may lose too much energy due to inelastic collisions with solvent molecules, thereby reducing the efficiency of the ICEC process.

The distance between the electron and the cation, as well as the electron's kinetic energy as a function of time, were analyzed to elucidate the dynamics of the ICEC process. We observed that the electron's kinetic energy decreases rapidly due to interactions with the solvent, and ICEC is most likely to occur at short distances and higher energies. After a few femtoseconds, the electron's energy often drops below the threshold required for ICEC, rendering the process energetically unfeasible.

The ICEC electron energy distributions at different times and concentrations were also examined. At shorter times, the electron retains higher energies, corresponding to trajectories where the electron reaches the cation quickly. At longer times, the electron's energy decreases, reflecting the energy loss due to interactions with the solvent. This behavior is consistent across different initial electron energies, with higher initial energies resulting in relatively faster ICEC events.

The semiclassical model presented here provides valuable insights into the dynamics of ICEC in aqueous solutions. It highlights the importance of considering both the initial kinetic energy of the electron and the concentration of cations in understanding the efficiency of the ICEC process. Future work could involve refining the model to include more detailed quantum mechanical effects, exploring different solvent environments, and investigating the role of other cations and anions in the ICEC process.

\section{Acknowledgments}
N.S. acknowledges the Agence Nationale de la Recherche (ANR) and the Deutsche Forschungsgemeinschaft (DFG) for their financial support through the QD4ICEC project (Grant No. ANR-22-CE92-0071-01). N.S. also expresses gratitude to the Indo-French Centre for the Promotion of Advanced Research (CEFIPRA) for funding under Project 6704-2. Additionally, N.S. thanks Prof. Lorenz Cederbaum for insightful and fruitful discussions.

\section{References}
\bibliography{refs.bib}

\end{document}